# Graphene-based magnetoelastic biosensor for COVID-19 serodiagnosis

Wenderson R. F. Silva[1,*], Larissa C. P. Monteiro[2], Murilo C. Costa[1], Renato V. A. Boaventura[3], Eduardo N. D. de Araújo[1], Rafael O. R. R. Cunha[1], Tiago A. de O. Mendes[2], Rodrigo G. Lacerda[3], Joaquim B. S. Mendes[1,†]

[1]Departamento de Física, Universidade Federal de Viçosa, 36570-90, Viçosa, Minas Gerais, Brazil

[2]Departamento de Bioquímica e Biologia Molecular, Universidade Federal de Viçosa, 36570-900, Viçosa, Minas Gerais, Brazil

[3]Departamento de Física, Universidade Federal de Minas Gerais, 31270-901 Belo Horizonte, Minas Gerais, Brazil

## ABSTRACT

This work presents an innovative magnetoelastic (ME) biosensor using graphene functionalized with the SARS-CoV-2 N protein for antibody detection via magnetoelastic resonance. Graphene was chosen for its biocompatibility and high surface area, enabling efficient antigen adsorption, validated by techniques such as EDX, AFM, and micro-Raman. Changes in Raman bands (a ~10 cm$^{-1}$ shift in the 2D band and an increase in the $I_D/I_G$ ratio from 0.03 to 0.60) confirmed non-covalent interactions and enhanced surface coverage with 100 µg of N protein. Tests using human plasma (10 RT-PCR-positive and 10 negative samples) demonstrated clear distinction between groups using graphene sensors functionalized with 100 µg of N protein. ELISA validation corroborated the results. Optimization of protein concentration and biofunctionalization time highlighted the importance of homogeneous surface coverage for reproducibility of the graphene-based ME biosensor. The platform combines graphene's advantages with the wireless, real-time detection capabilities of ME sensors, offering low cost, high sensitivity, and potential for automation, with applications in point-of-care diagnostics.

Corresponding authors: [*] wenderson.f@ufv.br, [†] joaquim.mendes@ufv.br

**INTRODUCTION**

The COVID-19 pandemic highlighted the need for fast, accurate, and accessible methods for diagnosing infectious diseases, a need that has been innovatively addressed by biosensors[1–5,19]. One material that has shown promise for biosensor platforms is graphene, which has a two-dimensional structure and unique properties such as high electrical and thermal conductivity, large surface area, and biocompatibility[6–10]. Its ability to efficiently adsorb biomolecules and promote specific interactions ensures good device sensitivity, enabling precise detection. Furthermore, the versatility of graphene in functionalization processes makes it particularly attractive for health applications, especially those based on interactions with antibodies and antigens[7–10].

Recent studies have demonstrated the potential of graphene-based biosensors in different configurations. Seo et al. (2020) developed a field-effect transistor biosensor functionalized with SARS-CoV-2 proteins, achieving high sensitivity in antigen detection. Timilsina et al. (2023) and Yakoh et al. (2021) developed different electrochemical biosensors for the quantification of SARS-CoV-2 and COVID-19 diagnosis, respectively, both based on graphene. These advancements reinforce the versatility of graphene in diagnostic platforms, however, the use of graphene in magnetoelastic (ME) sensors has not yet been reported. ME biosensors, due to their good sensitivity in detection, have proven to be an important platform for monitoring biomolecules and detecting changes in biological systems with high precision and sensitivity[3,11–14]. Moreover, they stand out for their ability to operate in low-cost, wireless, and real-time conditions[11].

In this sense, our work proposes an innovative ME biosensor that uses graphene as a base for antigen functionalization, enabling the detection of specific SARS-CoV-2 antibodies in human plasma. The proposal aims to fill a gap in current approaches, offering a low-cost, highly sensitive platform with great potential for automation. The integration of graphene with ME sensors opens new possibilities for applications in medical diagnostics, disease monitoring, and broader healthcare applications[1,2,5]. In addition to contributing to the advancement of biomedical diagnostics, this sensor combines the efficiency of the magnetoelastic platform with the unique properties of graphene, creating new opportunities for large-scale applications, such as epidemiological monitoring and point-of-care diagnostics[5].

## MATERIALS AND METHODS

*Obtaining, heterologous expression and purification of the recombinant protein N*

The recombinant protein N was obtained as previously described in Silva et al. (2024). Briefly, the optimized recombinant N-nucleocapsid phosphoprotein (NCBI sequence Gene ID: 43740575) with N-terminal His-tag and gene sequence with codon optimization for expression in *Escherichia coli* DH5α was commercially acquired and inserted into the pET28a expression vector (Biomatik, Canada). Recombinant plasmid transformation was performed by heat shock, and then the transformants were selected, the plasmids were extracted and verified by colony PCR, restriction enzyme digestion and sequencing. The confirmed recombinant plasmids were transformed into the *E. coli* Artics express strain for optimized recombinant N-nucleocapsid phosphoprotein expression in Luria Bertani (LB) broth supplemented with kanamycin, and the protein expression was then induced by adding 1 mM isopropyl-β-d-thiogalactopyranoside (IPTG). The bacterial culture pellets formed after centrifugation were lysed by sonication under denaturing conditions using the binding buffer composed of 20 mM imidazole, 20 mM NaH2PO4, and 0.5 M NaCl. After new centrifugation, the insoluble fraction was then collected, solubilized in solubilization buffer (binding buffer added 8 M urea), and applied to Ni-NTA columns (Qiagen, Hilden, Germany) in ¨AKTA-Start™ System (GE Healthcare Life Sciences) for purification of His-tagged protein using binding and elution buffer (20 mM $NaH_2PO_4$, 0.5 mM NaCl and 500 mM imidazole). The purified recombinant N-nucleocapsid phosphoprotein was analyzed by SDS-PAGE electrophoresis and Western blotting to evaluate their identity and purity.

*Plasma samples and ELISA*

The plasma samples used in this study belong to the plasma library of the Biotechnology and Molecular Biology Laboratory at the Federal University of Viçosa. The heat inactivated plasma samples were obtained in October 2020 during the COVID-19 pandemic, from patients admitted to the Materdei and Risoleta Tolentino Neve Hospitals (Belo Horizonte, Minas Gerais) and are registered in the Human Research Ethical Commission in Rene Rachou Institute (CAAE 30399620.0.0000.5091 - registration 4.210.316). In this work, ten plasma samples were collected from patients infected with the SARS-CoV-2 virus confirmed by the real-time reverse transcription PCR (RT-PCR) assay using nasopharyngeal swabs, and ten other plasma samples from negative patients were used. The plasma samples were firstly tested by the gold standard method enzyme-linked immunosorbent assay (ELISA) using the recombinant N-nucleocapsid phosphoprotein as antigen, for comparison of the biosensor serodiagnostic capacity (Silva et al., 2024).

*Surface biofunctionalization validation*

To confirm the immobilization of the recombinant antigen on the surface of the biosensor, energy dispersive X-ray spectroscopy (EDX), atomic force microscopy (AFM), and micro-Raman spectroscopy analyses were used to evaluate the functionalization steps. EDX spectra for elemental analysis of the biosensor surface before and after protein N immobilization were measured, using an accelerating voltage of 7 kV, both measures using the JEOL-JSM-6010LA microscope (JEOL Corporation, Tokyo, Japan). AFM analysis was performed to examine the immobilization effect of the antigen on the biosensor using the NT-MDT Integra Prima scanning probe microscope (NT-MDT, Zelenograd, Russia) at 23 ºC and 30% relative humidity operated in the tapping mode, obtaining images of 2 × 2 μm² area at a resolution of 256 × 256 pixels. The micro-Raman spectroscopy measurements were performed in an InVia Renishaw spectrometer (Renishaw, Watton-Under-Edge, United Kingdom) using a 785 nm and 514 nm excitation laser line focused by a 50 × objective lens. All measurements were collected with 15 s acquisition time and 4 spectral accumulations. The spectra were deconvoluted using Fityk software.

*Principle of detection and resonance frequency measurements of ME biossensors*

ME resonance occurs due to the strong coupling between the mechanical and magnetic energies in the material, leading to a variation in the elasticity modulus $E$ as a function of the applied magnetic field $H$. The relationship between the magnetic field $H$ and the elastic properties of the ME biosensor is given by the delta $E$ effect, in which the phenomenon of magnetostriction modifies the elasticity modulus $E$ of the ME material in the presence of the field $H$, as described by equation (1):

$$E(H, \sigma) = \left( \frac{1}{E_S} + \frac{9\lambda_s^2 H^2}{\mu_0 M_s H_{A\sigma}^3} \right)^{-1}, \quad (1)$$

where $E_S$ is the elasticity modulus without a magnetic field ($H = 0$), $\mu_0$ is the magnetic permeability of free space, $\lambda_s$ is the magnetostriction constant, $M_s$ is the saturation magnetization, and $H_{A\sigma}$ is the magnetoelastic anisotropy field dependent on the mechanical stress $\sigma$. For a biosensor vibrating freely in the air, the ME resonance frequency is given by equation (2):

$$f_r = \frac{n}{2L} \sqrt{\frac{E}{\rho(1 - v^2)}}, \quad (2)$$

where $n$ is the resonance mode ($n = 1$ for the fundamental mode), $L$ is the length of the biosensor, $\rho$ is the density, and $v$ is the Poisson's ratio.

The introduction of an external magnetic field modifies the relationship between stress and strain in the biosensor, altering its mechanical stiffness. When $H = H_{A\sigma}$, the sensor acquires larger vibration amplitudes, leading to increased precision and sensitivity to environmental disturbance conditions[11]. The deposition of an additional material layer on the sensor's surface alters the effective density and stress conditions in the ME material, directly impacting the elasticity modulus $E(H)$ and shifting the resonance frequency $f_r$. The perturbation in the ME biosensor is performed with magnetic fields consisting of a DC bias and an AC modulation component of frequency ω ($H = H_{DC} + h_{AC}(\omega)$), where the intensity of the $h_{AC}$ component is much smaller than $H_{DC}$. Since $f_r$ is related to the propagation of mechanical waves in the ME biosensor, any variation in stress or effective density due to the mass deposition on the surface of the ME biosensor results in a change in the vibrational response, enabling the precise detection and quantification of these changes.

For the excitation and detection of ME signals in the biosensors, a DC polarization coil and a pick-up coil were developed. The polarization coil, responsible for generating the DC bias component, was designed with 600 turns, a 23 mm inner diameter, and a 50 mm length, using 0.40 mm thick copper wire with a resistance of 10.7 Ω. This coil produces the $H_{DC}$ component of the magnetic field, establishing a static magnetic field that polarizes the sensitive material of the ME sensor. The coil was connected to a DC power supply from Keysight U8030 (Fig. 1c). For the pick-up coil, responsible for sweeping (generating the $h_{AC}(\omega)$ field) and detecting the AC signal, a coil with 200 turns, a 4 mm inner diameter, and a 22 mm length was developed, using 0.18 mm thick copper wire with a resistance of 4.5 Ω. This coil was positioned concentrically with the polarization coil and connected to a vector network analyzer (VNA) (R&SZNLE18, Rohde & Schwarz, Munich, Germany), operating in the $S_{11}$ mode with a sweep step of 5 Hz, for excitation and reading of the $f_r$ signal (Fig. 1a). The coil configuration was carefully chosen to maintain an appropriate number of turns, minimizing the increase in resistance and consequently the thermal noise, which is crucial to preserving the signal-to-noise ratio. All $f_r$ measurements were performed at room temperature (25 °C), with the sensor placed inside the pick-up coil, as shown in Fig. 1b.

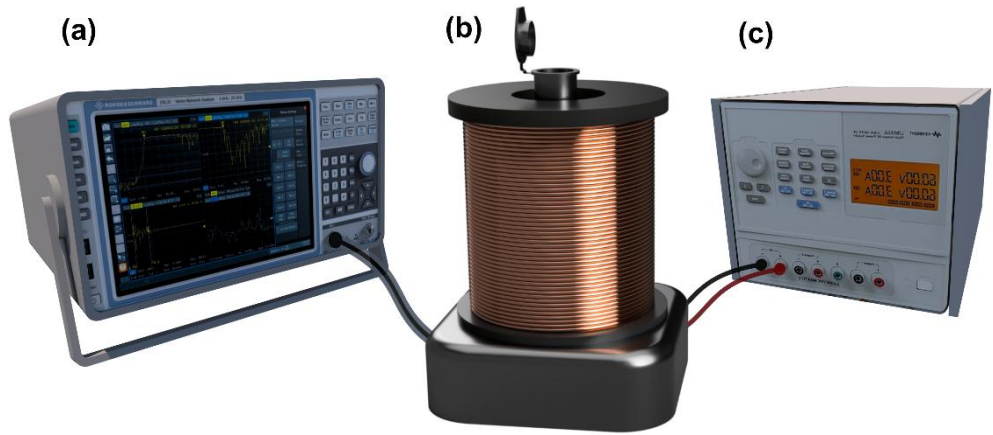

**Figure 1 -** (a) VNA connected to the pick-up coil (b). The external DC polarization coil, concentric with the pick-up coil, is connected to the Keysight DC power supply (c) for powering.

*ME Biossensor Preparation*

The ME biosensors were fabricated from METGLAS® 2826 MB3 alloy strips ($Fe_{40}Ni_{38}Mo_4B_{18}$), with dimensions of 5 mm × 1 mm × 28 μm. Cleaning was performed in an ultrasonic bath with acetone and then ethanol for 10 minutes at each step. After cleaning, a 50 nm layer of silicon dioxide ($SiO_2$) was deposited using RF sputtering. Finally, the biosensors were annealed at 200 °C for 2 hours in a vacuum oven (~$10^{-3}$ Torr) to relieve internal stresses and optimize the adhesion of the $SiO_2$ film. The $SiO_2$ deposition ensures better adhesion of the graphene layer to the ME strips, as well as providing a surface with high electronic affinity to graphene, eliminating possible contaminants that could affect the stability of the biomolecules deposited on its surface (BAGHERZADEH et al., 2015).

*Transferência do grafeno CVD and functionalization of ME biossensor*

Transferring graphene to a magnetoelastic substrate is crucial for exploring new applications in these platforms. Fig. 2 illustrates the transfer processes. Graphene was produced using the CVD (Chemical Vapor Deposition) method, a widely used technique for large-scale graphene production. The graphene was grown on Cu sheets inside a CVD chamber at 1000°C, with a gas mixture of $CH_4$ (33% by volume) and $H_2$ (66% by volume) at 330 mTorr for 2.5 hours. In step (1) of Fig. 2, a 120 nm thick PMMA layer is applied to the top surface of the Cu/Graphene to protect it from potential contamination. In the following step (2), the Cu is removed using an ammonium thiosulfate solution $(NH_4)_2S_2O_2$ (MENDES et al., 2019; SILVESTRE et al., 2015; BARCELOS et al., 2014). In step (3), the Graphene/PMMA film is washed in deionized water. The Graphene/PMMA structure is then transferred onto the Metglas, as illustrated in step (4), resulting in a Metglas/SiO$_2$/Graphene/PMMA junction, as shown in step (5).

The functionalization process initially involves the removal of PMMA by immersing the sensors in an acetone bath for 1 hour, followed by rinsing with isopropyl alcohol (both steps in non-agitated solution) and drying under a gentle stream of N$_2$ gas (SEO et al., 2020; SUK et al., 2011). The graphene surface of the ME biosensor was biofunctionalized with 2 µg and 100 µg of recombinant N-nucleocapsid phosphoprotein (Fig. 2, step (7)). The biosensor was incubated in solution for 1 hour at room temperature (25ºC). Afterward, the biosensor was rinsed with distilled water and dried on absorbent paper. Once biofunctionalized, the biosensors were exposed to plasma from patients confirmed or not for COVID-19 by RT-PCR, diluted in PBST at the same concentration used in the ELISA assay.

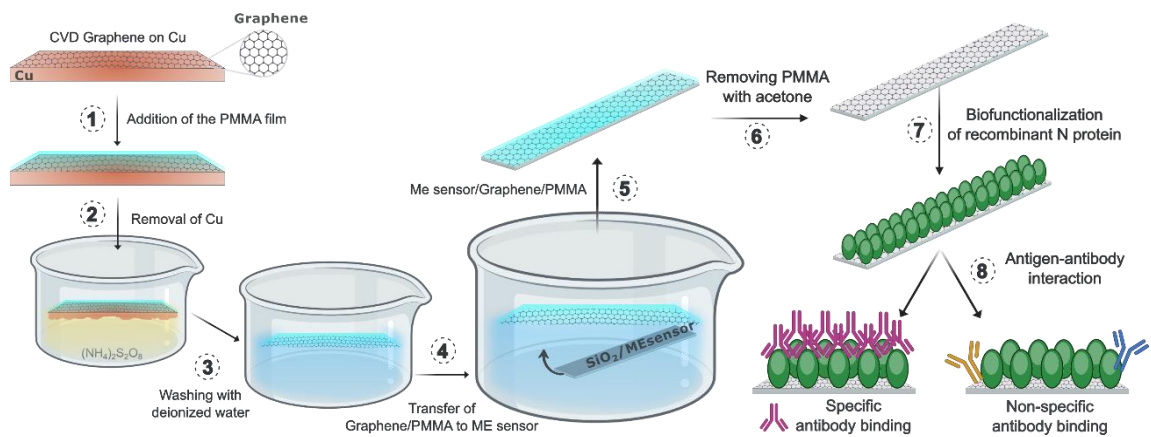

**Figure 2** - Schematic representation of the graphene transfer process used in this work. Graphene is produced on a copper plate (Cu (111)) and coated with PMMA for protection. After the Cu layer is removed, the film is transferred onto the surface of the ME sensor, which has a 50 nm layer of SiO$_2$. The PMMA removal process is carried out prior to biofunctionalization, as described in steps (7).

## RESULTS AND DISCUSSION

*Assessment of the surface functionalization of graphene*

Graphene, with its single-layer carbon structure and two-dimensional nature, exhibits unique properties such as high electrical and thermal conductivity, good biocompatibility, and a large surface area that facilitates the adsorption of functionalizing molecules (ASIF, 2015). These characteristics make graphene a promising material for the functionalization of biomolecules in biosensors, enabling a high density of biomolecule immobilization and potentially increasing the sensitivity of the ME biosensor. Moreover, its biocompatibility allows for interactions with biomolecules without compromising their activity or functionality.

The image in Fig. 3a shows the Raman spectrum of N protein in solution. In 3b, the spectrum of the protein on the ME biosensor with an active graphene surface is presented. The shift in the bands associated with N protein indicates that it interacts non-covalently with graphene, which is commonly observed in functionalization of this material with biomolecules (Zhan, 2022). In 3c, the EDX spectrum of the biosensor surface after the biofunctionalization process shows peaks associated with N protein, with the elements carbon (C), nitrogen (N), and oxygen (O) present elements found in the macromolecule of N protein as well as peaks associated with Si, Fe, and Ni, which are present in the ME biosensor. Based on the analysis presented in Figs. 3a-c, it can be concluded that the N-nucleocapsid phosphoprotein was successfully immobilized on the graphene surface of the biosensor.

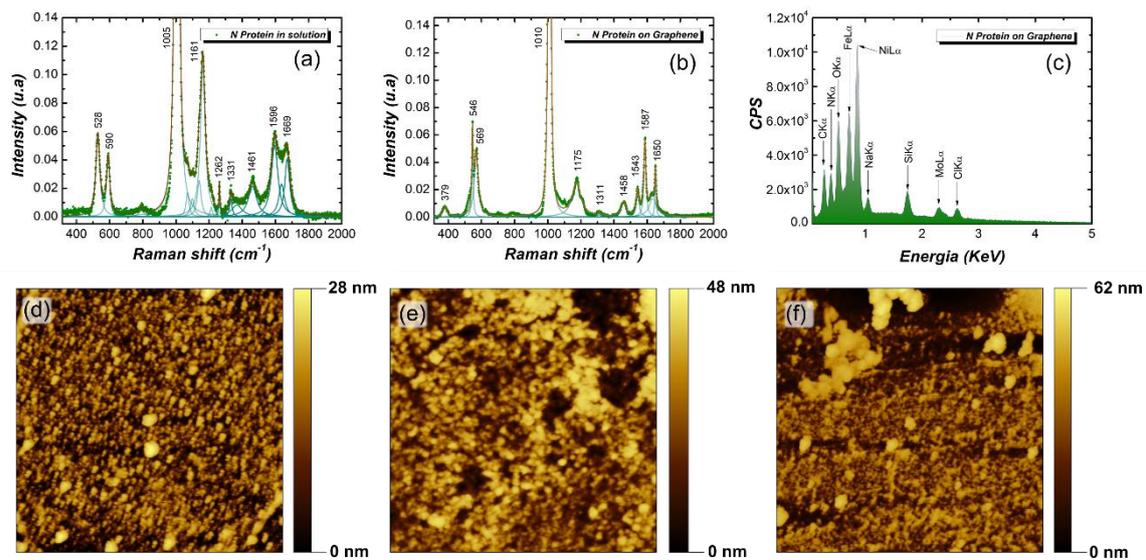

**Figure 3** - (a) Raman spectrum under excitation with 785 nm laser line of the recombinant N-nucleocapsid phosphoprotein in solution, and (b) Raman spectrum of the recombinant N-nucleocapsid phosphoprotein on the graphene surface of the biosensor after the functionalization process. (c) EDS spectrum of the recombinant N-nucleocapsid phosphoprotein on the graphene surface of the biosensor after the functionalization process. AFM images of the functionalized ME biosensor surface, captured from the biosensor with the blocking protein (d), after interaction with the negative serum (e) and positive serum (f).

The images in Fig. 3d-f present AFM images of the graphene surface of the ME biosensor coated with the N-nucleocapsid phosphoprotein + blocking protein BSA 3d, after interaction with negative 3e and positive 3f human samples, showing an increase in the relative height of the deposited material. The formed layer indicates the deposition of COVID-19 antibodies selected by the N-nucleocapsid phosphoprotein. Overall, the AFM images demonstrate that the functionalization of the ME biosensor surface with specific interaction with positive human samples was successful.

*ME biosensors for anti-Sars-Cov-2 antibody detection*

The results obtained from the measurements of the change in magnetoelastic resonance frequency are presented in Fig. 4. In 4a, the resonance frequency during the functionalization time for negative serum patients, and in 4b, for positive serum patients. In 4c, the shift in resonance frequency for positive and negative serum patients is shown over the exposure process. A distinction can be made between positive and negative serum based on the shift $\Delta f$. The incomplete saturation observed in 4c for the frequency shift $\Delta f$ may indicate that the functionalization time for graphene could be longer than that obtained for Au. Alternatively, the concentration used in the biofunctionalization process of N protein (in this case, 2 µg) may not be sufficient to achieve complete saturation of the active area of graphene with N protein, which could have resulted in a longer exposure time for saturation to occur. In this regard, a concentration of 100 µg of N protein was also tested in the biofunctionalization process to better optimize the performance of the functionalized graphene biosensor.

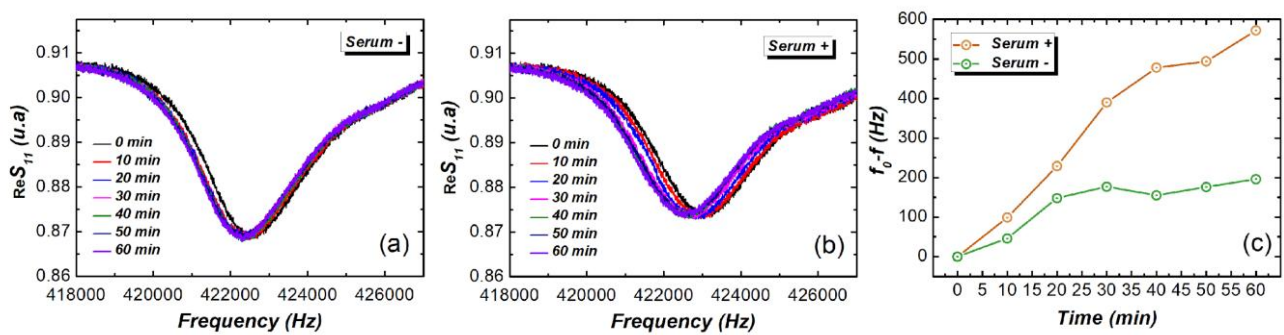

**Figure 4** - Magnetoelastic resonance spectra of the biosensor exposed to negative serum (a) and positive serum (b) during the biofunctionalization time. In (c), changes in the resonance frequency of the sensor exposed to positive and negative serum.

The frequency shifts $\Delta f$ for 20 biosensors using different concentrations and materials as active surfaces are shown in Fig. 5. The dispersion observed among the graphene biosensors (Fig. 5b) was greater than that obtained for Au (Fig. 5a), both biofunctionalized with 2 µg of N protein[3]. This result may be related to the concentration of N protein used for graphene functionalization, which can influence the uniformity and effectiveness of the formed layer, affecting the accuracy and reproducibility of the $\Delta f$ shifts in the biosensors[20]. In this regard, increasing the N protein concentration in the biofunctionalization process to 100 µg yielded better results (Fig. 5c). With this concentration, the biosensors were to fully distinguish seropositive from seronegative patients, indicating that the increased concentration in the biofunctionalization process led to lower dispersion in the obtained results. This improvement may be associated with greater coverage of the active graphene layer with N protein, enhancing the effectiveness of the biofunctionalization process. In the ELISA assay, positive samples showed an optical density of 2.00 ± 0.12, while negative samples exhibited 0.02 ± 0.07, ensuring 100% sensitivity, specificity, and accuracy in distinguishing between the groups.

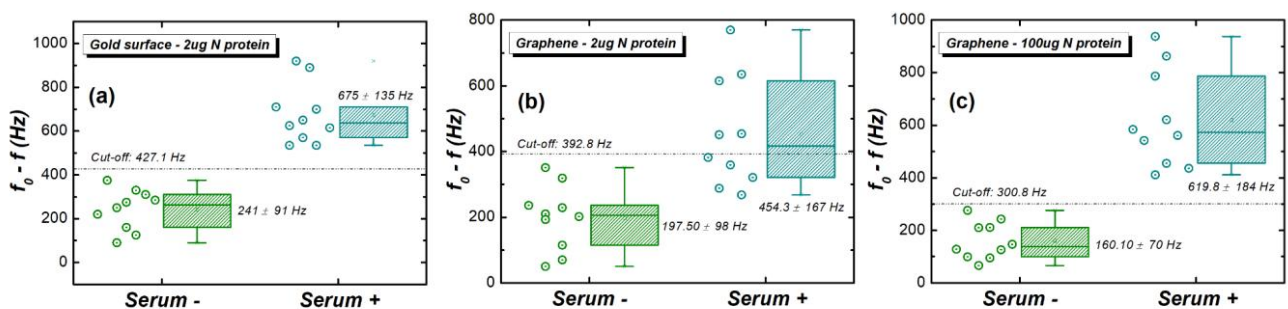

**Figure 5** - Resonance frequency shifts measured with the mean value and standard deviation (SD) for the twenty biosensors after exposure to positive plasma (blue) and negative plasma (green) for: (a) sensor with a gold surface biofunctionalized with 2 µg of N protein, (b) graphene surface biofunctionalized with 2 µg of N protein, and (c) graphene surface biofunctionalized with 100 µg of N protein.

In order to verify the interaction of N protein with graphene, Raman spectroscopy investigations were conducted varying the protein concentration used in the graphene surface functionalization, shown in Fig. 6. The sensitivity of micro-Raman spectroscopy to detecting small quantities of molecules makes it a suitable technique for measuring the interaction effects of biomolecules with monolayer graphene (Wu, 2024). The most prominent features of the Raman spectrum of graphene are the G band, associated with the in-plane stretching of C–C bonds, the D band, related to intervalley disorder-induced double-resonance scattering processes, and its overtone, the 2D band, which is an allowed Raman peak of graphene (Jorio, 2010). The 2D Raman peak shifts up by approximately 10 cm$^{-1}$ from pristine graphene to a surface functionalized with 100 µg, indicating p-doping of graphene (Das, 2008), which is consistent with the π–π interaction between the N-protein and graphene (Seo et al., 2020). The increase in the $I_D/I_G$ ratio from 0.03 to 0.60 suggests a higher density of defects, which can be attributed to increased surface coverage at higher N-protein concentrations. This could be related to the clear separation between positive and negative serum observed in the graph in Fig. 5c.

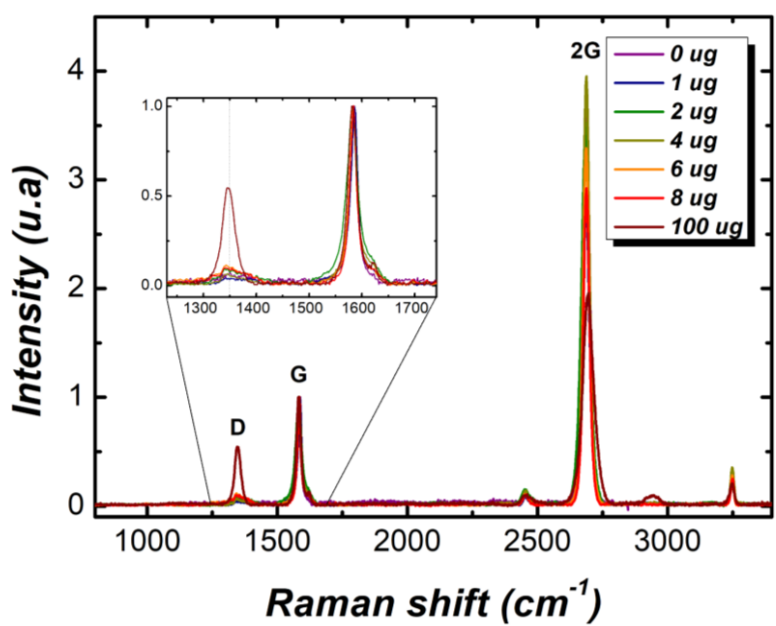

**Figure 6** - Raman shift for the graphene surface on SiO$_2$ biofunctionalized with different concentrations of N protein, normalized by the intensity of G Raman peak.

## CONCLUSIONS

This study demonstrates the efficacy of graphene-coated magnetoelastic (ME) biosensors for detecting COVID-19 antibodies through biofunctionalization with the N protein. Micro-Raman and EDX measurements confirmed robust protein adhesion, while AFM revealed distinct surface modifications in biosensors exposed to positive human sera compared to negative controls. ME resonance measurements showed significant frequency shifts exclusively in SARS-CoV-2-positive plasma, enabling optimization of biofunctionalization time and antigen concentration. Micro-Raman spectroscopy further validated non-covalent interactions between graphene and the N protein, evidenced by 2D band shifts (~10 cm$^{-1}$) and increased $I_D/I_G$ ratios. Combined structural analyses (EDX, AFM, Raman) and functional assays (resonance frequency, ELISA) underscore the biosensors precision and reproducibility. The integration of graphene's biocompatibility with ME technology offers a cost-effective, sensitive, and automatable platform for COVID-19 diagnostics, highlighting its potential for epidemiological surveillance and point-of-care applications.

## ACKNOWLEDGMENT

This research is supported by the Conselho Nacional de Desenvolvimento Científico e Tecnológico (CNPq), Coordenação de Aperfeiçoamento de Pessoal de Nível Superior (CAPES), Financiadora de Estudos e Projetos (FINEP), Fundação de Amparo à Pesquisa do Estado de Minas Gerais (FAPEMIG) — Rede de Pesquisa em Materiais 2D and Rede de Nanomagnetismo — and the INCT of Spintronics and Advanced Magnetic Nanostructures (INCT-SpinNanoMag; CNPq Grant No. 406836/2022-1).